\documentclass[11pt,dvips]{article}
\usepackage{graphicx}
\begin{document}
\begin{center}
{\Large Turbulent intermittency and Euler similarity solutions\\}
{Daniel P. Lathrop\\
Department of Physics\\
Institute for Research in Electronics and Applied Physics\\
Institute for Physical Sciences and Technology\\
University of Maryland, College Park, MD 20742}
\end{center}

{
Self-similar Euler singularities may be useful for
understanding some aspects of
Navier-Stokes turbulence.  Here, a causal explanation
for intermittency is given, based on the control of 
the sudden growth of the gradients
by the Euler equations.  
This explanation uses certain Euler solutions as 
intermediate asymptotics in Navier-Stokes
turbulence \cite{barenblatt} -- controlling the
dynamics over a limited spatial and temporal domain.
These arise from an analysis of similarity equations,
previously discussed by Pelz and Green \cite{pelz},
which yield experimentally testable predictions.   
Three main points are presented here: scalings of suitable
characteristic lengths with time
from a critical time $l \sim
(t_\circ -t)^\alpha$, $\alpha>1$, a discussion of invariant sets 
of the similarity equations
that result, and a discussion of cutoff mechanisms.  The value $\alpha = 3/2$
appears to
correspond to Kolmogorov scaling for turbulence.
Some limited experimental evidence is presented from Eulerian gradient
measurements at the Kolmogorov scale showing $1<\alpha<3$ values. 
Much testing is necessary to 
ascertain the final usefulness and validity of these ideas, as several
conceptual obstacles remain.
}
\vspace{0.30 in}

\section{Introduction}

Turbulent flows of liquids and gases occur in a 
host of natural and engineering phenomena.  A detailed understanding of 
atmospheric flows on Earth and the planets, predictive design for
industrial equipment, and understanding of the processes within the liquid
cores of planets all rely on an understanding of turbulence.
One might naively think this an easy task, given that the governing equations 
are well known (i.e., the family of equations stemming from the
Navier-Stokes equations with appropriate body forces).  
That understanding is hampered by the inherent nature of turbulence itself,
where the velocity field is
rough and velocity gradients are intermittent.  
Phenomenological considerations of Richardson and
Kolmogorov have been very fruitful \cite{K41, K41b}, but have
also left deep holes in our understanding of the intermittent 
nature of turbulence\cite{frisch}.
By intermittency, I mean a host of interlocking observations that 
the basic observables in 
turbulent flow; the vorticity, dissipation, helicity, 
acceleration, and their ability
to effect advected scalar and vector fields 
(concentration or perhaps magnetic fields)
all show abnormal statistics.  As one example, in experiments one can observe 
accelerations hundreds times the means, lying 30 standard deviations from those means
\cite{bodenschatz,pinton}, or observe the dissipation locally becoming one hundred times its
average value \cite{meneveau, zeff, sreeni2}.  The probability distributions for these
observable then show far from Gaussian statistics with long 
tails.  These types of gradient or time derivative concentrations appear local
in space and time.  Recent results of Gibbon and Doering suggest that
the time intervals where singularities are possible become more localized
as the Reynolds number increases \cite{gibbon}. 

There is little in the phenomenological picture of 
a cascade of energy to smaller scale
(envisioned in wavenumber space) to explain intense events.  The framework
discussed here is an attempt to explain 
these observations, although it is as yet a hypothetical explanation.  
Rather than attempting a phenomenological explanation,
instead I seek a causal explanation rooted in the 
known equations of motion and the relevant
forces.   

This is a non-statistical approach, but with the need 
to ultimately explain the observed statistics.
First, I argue that the viscous forces are not responsible for the
sudden growth locally of gradients, and that in this context, 
during the growth, one should consider
the viscous forces irrelevant.  This limits our scope to the growth phase
of these events, and lands us in the context of the Euler equations.
Solutions to the Euler equations may explain 
the sudden growth in the gradients
and time derivatives 
in the context of an intermediate asymptotic \cite{barenblatt}.  
These solutions will be locally
relevant for only a range in times, nearby the sudden collapse, but limited to times before viscosity,
or perhaps cavitation, act to limit the strength of these near-singularities.  
As an example, one can look to the Burgers equation \cite{burgers}.  
There, negative slopes tend to collapse
to form shocks, but the gradients are limited by the action of viscosity.  In the case when the viscosity
is quite small, there is a substantial range in time where the gradient 
value at the location of minimum
gradient is approximately governed by 
the inviscid Burgers equations, with the gradients 
growing as predicted, i.e. inverse
in time from a critical time.

What is unlikely to be understood through this 
exercise are the important long lived vortices so many have
remarked on \cite{vortex}, which form an independent, but perhaps 
causally connected part of turbulent phenomena.
Remarks on the connection between long lived vortex filaments 
and the local collapse of gradients
studied here are given in the conclusion.

When entering the study of gradient focusing and local intermittent events in turbulence, one treads
on the long worn and active path of singularities in Navier-Stokes and Euler equations \cite{cantwell, pelz2, kerr, tanveer, eggers2, siggia, 
pelz3, pumir, brachet, ohkitani}.  
Although these
studies are conceptually related, I view the Euler solutions discussed here (which are singularities of
the Euler equations) as merely intermediate asymptotics for Navier-Stokes, as remarked earlier, and not
observable singularities of viscous fluid flow.  
Many previous studies have concentrated on the Leray scaling $\alpha=1/2$ \cite{leray, pelz} or
on the case where a co-collapsing sphere shows energy conservation $\alpha=2/5$ \cite{constantin}.  At least one simulation of R.M. Kerr has shown some
evidence for $\alpha=1$ behavior \cite{kerr}.
I restrict my attention to the case (described later) of $\alpha>1$ so that, as observed in turbulence,
the velocities are not growing abnormally large.  In many turbulent flows the probability
distribution of the velocity is close to Gaussian, while the gradients are intermittent.

Whether or not there are
singularities in Navier-Stokes solutions is perhaps of little practical consequence -- extra phenomena
such as dilations, cavitation or extreme rarity render them of little relevance.  One should not mistake the
irrelevance of singularities of the Navier-Stokes equations for an argument for the same for Euler
singularities.  
Here, I'm proceeding based on the premise 
that {\it the causal structure of Euler singularities is the underlying cause of
intermittency in Navier-Stokes turbulence}, even if they are capped by viscosity or other extra constraints.

While the following calculations and observations are presented with the 
hope of better understanding Navier-Stokes turbulence, their final utility 
is yet unknown.  Nevertheless, because of the possibility of their importance, 
it seems useful to present them. 

\section{Similarity methods}

There have been significant successes in understanding singularities 
and self-focusing in 
strongly nonlinear physical systems using similarity methods 
\cite{barenblatt,eggers,lister,lister2,lister3}.
Our previous work on the collapse of waves and the production of jets on a free surface illustrates
the utility of similarity techniques in allowing analytical and numerical progress on systems
not otherwise amenable to analysis \cite{zeff2000}.  While there is no guarantee of the success of this approach for Euler
collapse, proceeding with this hope may lead to new insight.  
Observe that in local gradient collapse the length
scales become small as one approaches maximum focusing.  
In what follows I will assume that useful information can be obtained by
consideration of the hypotheses that
the length scales in all directions behave
equally, and that the functional form is a power law in time. 
Regarding the first hypothesis, one might say that the
different directions may have trouble behaving in different manners as they
are constantly effecting each other and being mixed by motions and rotations.  This argument, however,
is hardly compelling.  One might also argue that the relevant equations 
balance linear and nonlinear terms,
so power-law time dependences may arise.
This, however, is not necessary and depends on the
algebraic form of these balances.  
There might be collapse events of more complicated forms;  
understanding their analytical structure may be much more difficult.  

Beginning with the Euler equations for incompressible flow:
\begin{equation}
\partial_t \vec{v} + \left(\vec{v} \cdot \vec{\nabla} \right)\vec{v} +
\frac{1}{\rho} \vec{\nabla}P=0\end{equation}
\begin{equation}\vec{\nabla} \cdot \vec{v} = 0  \mbox{,}
\end{equation}
one can apply the above assumptions to form an intelligent guess 
for the form of the solutions:
\begin{equation}
\vec{v}(\vec{x},t)=v^* \left(\frac{t_o - t}{t^*}\right)^{\alpha-1} \vec{G}
\left(\frac{\vec{x}}{L^*} \left(\frac{t^*}{t_o - t} \right)^\alpha\right)
\end{equation}
\begin{equation}
P(\vec{x},t)= \rho (v^*)^2 \left(\frac{t_o - t}{t^*}\right)^{2\alpha - 2} \Pi
\left(\frac{\vec{x}}{L^*} \left(\frac{t^*}{t_o - t} \right)^\alpha\right)
\end{equation}
Here various quantities are made dimensionless using some characteristic large scale $L^*$,
velocity scale $v^*$, and time scale $t^*=L^*/v^*$, as would be natural in the context of a shear flow.
One can then seek solutions for the time independent similarity 
functions $\vec{G}$ and $\Pi$.
The above ansatz assumes all lengths scale as $l \sim (t_o-t)^\alpha$, velocities as $v \sim (t_o-t)^{\alpha - 1}$. 
It is important to require that $\alpha \ge 1$ in order that the velocities remain bounded as we approach $t_o$ from below.
By defining a scaled coordinate $\vec{r}$ and a scaled gradient
\begin{equation}
\vec{r}=\frac{\vec{x}}{L^*}\left(\frac{t^*}{t_o - t}\right)^\alpha
\end{equation}
\begin{equation}
\hat{\nabla} = \left(\frac{t_\circ-t}{t^*}\right)^\alpha L^* \vec{\nabla}
\end{equation}
one can eliminate time dependence from the Euler equations and obtain what are referred to as similarity equations
\begin{equation}
(1-\alpha) \vec{G} +\alpha(\vec{r} \cdot \hat{\nabla})\vec{G} + \left(\vec{G} \cdot \hat{\nabla} \right)\vec{G} + \hat{\nabla}\Pi=0 \mbox{,}
\label{simeq}
\end{equation}
\begin{equation}
\hat{\nabla} \cdot \vec{G} = 0  \mbox{.}
\label{simin}
\end{equation}

Solutions to Eqs. \ref{simeq} and \ref{simin} correspond to solutions to
the original Euler equations.
Symmetries of the Euler equations may be used to generate further solutions 
associated with the similarity
ansatz.  The time or spatial origin of the solutions may be shifted, 
and one may also add a constant velocity. 

The solutions to Eq. \ref{simeq} are assumed to be smooth -- they represent solutions which retain their
smoothness up to the critical time $t_\circ$, 
and loose it by the length scales limiting to zero.  At that critical
time the solution remains smooth at all locations (governed by the far field of Eq. \ref{simeq}) except
at the origin, where the gradients become unbounded.  Pelz and Green 
\cite{pelz} have examined the
stability properties of these solutions.  Here I present new results and
discussion on the form of
the solutions of Eq. \ref{simeq}, their correspondence to Lagrangian dynamics, 
viscous cutoff mechanisms,
and their relationship to turbulence with Kolmogorov statistics.

I examine the similarity equations progressively in one, two and 
three dimensions, 
in order to better understand their function and utility.

\section{One dimensional similarity equations}

In one dimension, one neglects the requirement of incompressibility and drops
the pressure term in Eq. \ref{simeq}. 
In this case:
\begin{equation}
(1-\alpha) G + \alpha x \partial_x G + G \partial_x G = 0 \mbox{.}
\end{equation}
This equation has several families of implicit analytical solutions
\cite{pego}, only one of which is relevant here: 
\begin{equation}
x = -G + a G^{\alpha/(\alpha-1)}
\label{burg2}
\end{equation}
The far field (i.e., large $|x|$) 
limit of Eq. \ref{burg2} is $G=x^{1-1/\alpha}$.
This similarity solution gives the generic formation of the first shock in the inviscid Burgers equation, and accurately
describes the formation of a near-shock in the diffusive case, when the diffusion is small.  In particular, 
Burgers equation is unstable where the slope has a 
maximum negative value, the magnitude of the slope following a characteristic
diverges as $(t_\circ -t)^{-1}$ for a time until the diffusive cutoff takes over.
The value $\alpha=3/2$ is generically seen, due to a curious reason, and leading to a 
$G \sim |x-x_\circ|^{1/3}$ far field.  
The location for the characteristic 
which sees the first shock in the inviscid case is the most negative slope in the initial condition.  Generically, the
second derivative will vanish there; locally the curve will be dominated by 
the linear and cubic terms.  These cubic
terms yield, in the confluence of characteristics in the shock, a functional form at 
the time of the shock $u = u_\circ + v_* |(x-x_\circ)/L^*|^{1/3}$.
The probability distribution of negative gradients in randomly forced Burgers
turbulence is governed by these pre-shocks \cite{sinai}.

\section{Two dimensional similarity equations}
The situation in two dimensions is quite different.  Due to the simple advection of vorticity in two dimensions \cite{majda} 
-- the vorticity remains bounded and no similarity solutions exist.  
However, examining this in more detail 
will assist in understanding the three-dimensional case.

In two dimensions, the vorticity vector lies normal to the plain of motion. 
Due to this, there are no gradients of the vorticity along the direction
of the vorticity, and one can write the vorticity similarity equation (curl of Eq. \ref{simeq}) as:

\begin{equation} 
\omega + \alpha (\vec{r} \cdot \hat{\nabla}) \omega + (\vec{G} \cdot \hat{\nabla}) \omega = 0
\end{equation}

I can demonstrate that these equations have no solutions using several observations.
First, define the associated vector field $\vec{H} = \alpha \vec{r} + \vec{G}$.  
This field is used to rewrite the vorticity similarity equations in two dimensions as: 

\begin{equation} 
\omega + (\vec{H} \cdot \hat{\nabla}) \omega = 0
\label{vort2d}
\end{equation}

It is useful to define the dynamical system: 
\begin{equation}
\frac{d\vec{r}}{ds} = \vec{H}
\label{hflow}
\end{equation}
which
yields characteristics for the vorticity.  In terms of the trajectory $\vec{r}(s)$
one can write the vorticity equation along these characteristics as 
\begin{equation}
\frac{d \omega}{d s} = - \omega \mbox{.}
\end{equation}
The dynamical system for $\vec{r}(s)$ is everywhere expanding, as $\hat{\nabla}
\cdot \vec{H} = 2 \alpha$ and $\alpha$ is positive.  Nearly all trajectories
starting out near the origin escape to infinity.  Due to the (assumed) 
smoothness of the dynamical system,
Eq. \ref{hflow}, there must be an 
invariant set near the origin which does not escape.  As this calculation is 
restricted
to two dimensions, the possible invariant sets (by the Poincare-Bendixon
theorem \cite{guck}) may be fixed points or closed orbits.
Along any characteristic $\vec{r}(s)$ the vorticity must be $\omega(s) =
\omega(0) e^{-s}$.  At fixed points, by definition, $\vec{H}=0$, giving
$\omega = 0$ from Eq. \ref{vort2d} at those points.  For closed orbits, the vorticity $\omega$
must also be zero so that after going around one orbit it returns to its
original value.  Points starting nearby the closed orbits or fixed points
escape to large radius.  As their initial vorticity must be zero, as they
move out from near the origin their vorticity remains zero.  From this
one can understand that the vorticity is zero everywhere in the plain.  
The far field (large $|\vec{x}|$) solution for $\vec{G}$ must have a 
dependence rising like $r^{(\alpha -1)/\alpha}$, where the first
two terms of Eq. \ref{simeq} are dominant.  
As the vorticity is zero (i.e., $\hat{\nabla}\times \vec{G}=0$), 
the vector field $\vec{G} = \hat{\nabla} \phi$ must
be a gradient field with $\nabla ^2 \phi = 0$.  The far field dependence
is incompatible with a smooth Harmonic function $\phi$ which must have an
integer $n$ far field dependence $r^n$.   The far field exponent $(\alpha -1)/\alpha$ is not an integer for all cases of interest $\alpha > 1$. 
Thus, the only solution to the
similarity equations in two dimensions are $G=\omega=0$.
Another proof is also possible using integral methods \cite{tadmor}.

\section{Three dimensional similarity \\ equations - Lagrangian dynamics}

The flow generated by $\vec{H}$ is also of considerable use in understanding the
three dimensional case.  Trajectories of Eq. \ref{hflow} 
are directly associated with particle trajectories $\vec{x}(t)$
in lab coordinates.  The invariant sets are especially important
and correspond to trajectories which undergo the most singular Lagrangian motions.
Trajectories in the laboratory frame follow a path $d\vec{x}/dt = \vec{v}$;
this equation is equivalent to Eq. \ref{hflow} under the identifications
of $s = -\ln[(t_\circ -t)/t^*]$ and Eq. 5 for $\vec{r}$.
The correspondence between these different quantities is summarized in
the diagram:

\begin{equation}
\mbox{
\begin{tabular}[pos]{ccccc}
	&{\Large $t$} & {\Large $\stackrel{\vec{v}}{\rightarrow}$ }& {\Large$\vec{x}$ }& \\
{\small	$-\ln{[(t_{\circ}-t)/t^*]}$ }&{\Large $\downarrow $}& &
{\Large  $\downarrow$ }&{\small $\vec{x}[t^*/(t_{\circ}-t)]^\alpha/L^*$} \\
&{\Large $s$} &{\Large $\stackrel{\vec{H}}{\rightarrow}$ }& {\Large $\vec{r}$ }& \\
\end{tabular}
}
\end{equation}

Note as $(t_\circ - t) \rightarrow 0$, that $s \rightarrow \infty$.
As the flow generated by $\vec{H}$ has divergence everywhere positive ($\hat{\nabla} \cdot \vec{H} = 3 \alpha$),
trajectories all flow outward toward large radius, 
and there must be repelling invariant sets in the vicinity of the origin (assuming of course that
the similarity solution is smooth and exists!).  Several types of invariant sets are possible: fixed points, closed orbits,
or perhaps strange saddles \footnote{
2-tori are excluded, since their interior volumes would be invariant,
contradicting $\hat{\nabla}\cdot\vec{H}=3 \alpha >0$ which implies
that all volumes grow exponentially as $\exp(3\alpha s)$ following the
flow defined by Eq. \ref{hflow}.}.
Which type of invariant set will be determined by the precise form of the outer solution.
Assuming these solutions exist, and pending a numerical solution, it is useful
to explore the consequences of these invariant sets.
Fixed points of the dynamical system Eq. \ref{hflow} are defined by 
$\vec{H}=0$; we label them $\vec{r}^*$.
These points have a number of properties and yield testable prediction for the Lagrangian motion.
At points $\vec{r}^*$, one has that $\alpha \vec{r} = -\vec{G} = -\vec{\nabla}\Pi$ from the definition of $\vec{H}$ and
the similarity equation (Eq. \ref{simeq}) with $\vec{H}=0$.  

We can see why the non-existence proof for two dimensions fails in three by examining the vorticity
equation in three dimensions (the curl of the similarity Eq. \ref{simeq}):
\begin{equation}
\vec{\omega} + (\vec{H} \cdot \hat{\nabla}) \vec{\omega} = ( \vec{\omega} \cdot \hat{\nabla} ) \vec{G} \mbox{.}
\label{simvort}
\end{equation}
Rather than the vorticity being zero at the fixed points of Eq. \ref{hflow}, 
the vorticity must be a eigenvector of the gradient matrix with unit eigenvalue $M \vec{\omega} =
\vec{\omega}$.   
Although the 2-D non-existence does not carry over to 3-D, I 
have yet been unable to show existence directly,
numerics may give some evidence in future work.

From the equivolency of flows of $\vec{H}$ and Lagrangian trajectories, one can calculate the form for observables
due to the fixed points and limit cycles of the similarity flow.  For a fixed point $\vec{r}^*$ transformed to
laboratory coordinates the trajectory is
$\vec{x}= L^* [(t_{\circ}-t)/t^*]^\alpha \vec{r}^* + \vec{v_\circ} t + \vec{x_\circ}$, where
we have allowed for a background velocity and origin.  The Lagrangian velocity would be 
$\dot{\vec{x}}= - \alpha L^* [(t_\circ -t)/t^*]^{\alpha-1} \vec{r}^*/t^* + \vec{v_\circ}$; 
acceleration would be 
$\ddot{\vec{x}} = \alpha (\alpha-1) L^* (t_\circ-t)^{\alpha-2} \vec{r}^*/(t^*)^2$.  
This acceleration diverges at the critical time, clearly
in need of a cutoff due to viscosity or other physical intervention.  These forms, as intermediate asymptotics
are testable predictions for Lagrangian experiments and numerics -- are events observed which hold such forms?

In the case of an invariant cycle in the flow of $\vec{H}$ one obtains a different type of motion.  This rotating
motion in similarity coordinates, with some characteristic frequency in $s$,  yields an accelerating chirp of
spiraling motion.  For simplicity, let's say some some 
similarity coordinate has harmonic motion 
$r_1 = A \cos( \Omega s )$.   
Carried to laboratory coordinates the motion is
\begin{equation}
x_1 = A L^* (\frac{t_\circ - t}{t^*})^\alpha \cos[ \Omega \ln(\frac{t_\circ - t}{t^*}) ]\mbox{.}  
\end{equation}
The laboratory Lagrangian frequency thus rises as $(t_\circ-t)^{-1}$ while
the radius collapses to zero as $(t_\circ-t)^\alpha$.  Such motions may be related to the large accelerations
observed in recent Lagrangian tracking experiments \cite{bodenschatz, pinton}, and may account partially for
the extreme intermittency seen in the acceleration distributions.  It is a likelihood that a clear 
understanding
of the viscous cutoff is needed to fully account for such distributions;
some aspects of the viscous cutoff are discussed in Sec. 8. 
These calculations
give a plausible specific cause for extreme Lagrangian accelerations -- 
which is independent of the intense long-lived 
vortices which form an alternative explanation for those observations.

\section{Far field behavior}

The large $r$ behavior is particularly significant, as close to the
critical time, the near field character has shrunk (in laboratory coordinates)
into an insignificant volume, leaving a relatively large area displaying
the far field behavior in its wake.
For large $r$, one can calculate the behavior of three dimensional
solutions of Eq. \ref{simeq}.  Only the first and second terms are relevant, the
nonlinear term and pressure are sub-dominant.  In that case the resulting 
equation
$(1-\alpha) \vec{G} + \alpha (\vec{r} \cdot \hat{\nabla}) \vec{G} =0$ is
analytically solvable:
\begin{equation}
\vec{G} = r ^ {(\alpha-1)/\alpha} \vec{f}(\theta,\phi) 
\end{equation} 
for spherical coordinates $r$, $\theta$, $\phi$.  The function $\vec{f}$ is
a general $S^2 \rightarrow {\cal R}^3$ vector field subject only to $\hat{\nabla}
\cdot \vec{G} = 0$.

Thus observe that the far field radial dependence is fixed by $\alpha$, while
the angular dependence is general.  It has been conjectured \cite{laurette}
that Eq. \ref{simeq} has solutions for all admitable far field forms, while
it is important to
emphasize solutions (numerical or analytical) have not yet been found.

The far field behavior has particular significance if one hopes
to use these solutions to understand turbulence.  The far field solution
when $\alpha = 3/2$, if applied in many events, would lead to velocity
difference statistics associated with Kolmogorov scaling and the known
scaling for the third order structure function. 
When $\alpha = 3/2$ the far field behaves as $||\vec{G}|| \sim r ^{1/3}$, which
is retained in the lab frame velocity near the critical time.  This connection
in particular has been a motivation for seeking a connection between these
Euler similarity solutions and Navier-Stokes turbulence.  Why the $\alpha
= 3/2$ case may be prevalent is precisely unknown, but discussed in the
conclusion.

\section{Near field behavior}

One can learn about the structure of the gradients at the fixed points by examining the similarity
equation.  Taking the gradient of that equation, to obtain a matrix equation for $M = \partial_i G_j$, one obtains
at the fixed points $\vec{H}=0$ that:
\begin{equation} M + M^2 + {\Pi}_2 =0 \mbox{.}
\label{mateq}
\end{equation}

This matrix Riccati equation is related to the quadratic nonlinearities
in the Euler equations.  Gibbon \cite{gibbon2} has shown how quaternions
can be used in this context to relate Euler equations and the equations
of ideal magneto-hydrodynamics to a Schr\"{o}dinger equation.
Equation 20 yields nine constraints on 
the components of $M$ and $\Pi_2$.  Rotating $M$ into a form
where the symmetric part is diagonal, and given that $\Pi_2$ is symmetric, there are eleven independent components
between them.  Solving Eq. \ref{mateq} yields three types of solutions, each in two parameters.  Further solutions
come from permuting the axis in this form, or by rotations relaxing the diagonalization of the symmetric part
of $M$.

\begin{equation}
\begin{array}{l}
M = \left(
\begin{array}[pos]{ccc}
a & 0 & 0\\
0 & b & 0\\
0 & 0 & -a-b\\
\end{array}
\right) \\
\Pi_2 = \left(
\begin{array}[pos]{ccc}
a(a+1) & 0 & 0\\
0 & b(b+1) & 0\\
0 & 0 & (a+b)^2-(a+b)\\
\end{array}
\right) \\
\end{array}
\end{equation}

\begin{equation}
\begin{array}{l}
M = \left(
\begin{array}[pos]{ccc}
1 & 0 & 0\\
0 & a & f\\
0 & f & -1-a\\
\end{array}
\right) \\
\Pi_2 = \left(
\begin{array}[pos]{ccc}
2 & 0 & 0\\
0 & a+a^2-f^2 & 0\\
0 & 0 & a+a^2-f^2\\
\end{array}
\right) \\
\end{array}
\end{equation}

\begin{equation}
\begin{array}{l}
M = \left(
\begin{array}[pos]{ccc}
1 & 0 & f\\
0 & 1 & g\\
f & g & -2\\
\end{array}
\right) \\
\Pi_2 = \left(
\begin{array}[pos]{ccc}
2-f^2 & -fg & 0\\
-fg & 2-g^2 & 0\\
0 & 0 & 2-f^2-g^2\\
\end{array}
\right) \\

\end{array}
\end{equation}

The gradient matrix at these fixed points gives rise to rapidly growing
gradients in laboratory coordinates.  At these points, taking the gradient
of Eq. 3, yields:
\begin{equation}
\frac{\partial v_i}{\partial x_j} = M v^* \left(\frac{t_\circ -t}{t^*}\right)
^{-1}  
\end{equation}
leading to the vorticity and strains all growing jointly and suddenly.

\section{Viscous cutoff}

\begin{figure}[h]
	\begin{center}
		\includegraphics[width=3.5 in]{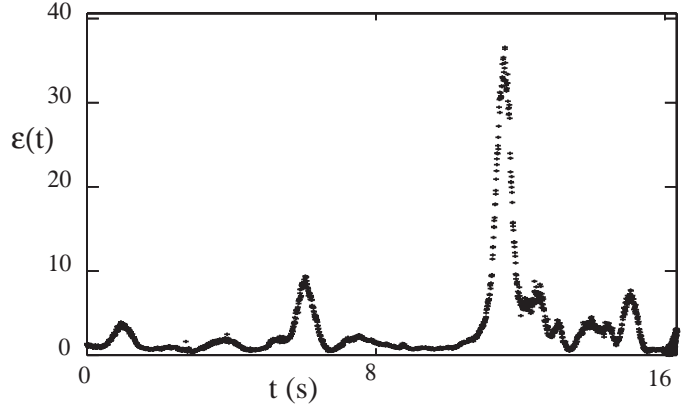}
	\end{center}
	\caption{Time trace of the dissipation measured at the Kolmogorov scale in an oscillating grid experiment. }
	\label{fig:fig1}
\end{figure}

One can obtain estimates for the small scale cutoff for these events by analyzing the
relevant forces.  For all events with $\alpha>1$ viscosity will be relevant at some late time,
where the length scales become small enough for the local Reynolds number to be of order unity.
Comparing the inertial forces with viscous forces for the scaling ansatz, $(\vec{u} \cdot \vec{\nabla})
\vec{u} < \nu \nabla^2 \vec{u}$ yields an expression for the time when 
(for smaller $(t_\circ - t)$) viscosity will be dominant.  This happens when 
\begin{equation}
\frac{t-t_\circ}{t^*} < (R^*)^{1/(1-2\alpha)}\mbox{,}
\end{equation}
where $R^* = u^* L^* / \nu$.  
This leads to a definition of a crossover time $t_k$ where the equality holds above.
At the crossover time, the length scale is
\begin{equation}
l_k = L^* (R^*)^{\alpha/(1-2\alpha)}\mbox{;}  
\end{equation}
this corresponds to the Kolmogorov scale for that event, and depends on
the value of $\alpha$.  In particular, events with $\alpha = 3/2$ have the normally defined Kolmogorov
scale, while events with $\alpha < 3/2$ have a reduced minimum scale.  
This expression is similar to those obtained for the Kolmogorov length in the multifractal
description of turbulence, where $l_k \sim R^{-1/(1-h)}$ for the Holder exponent $h$ \cite{bohr}.

This fluctuating Kolmogorov
scale effects the maximum gradient observed in different events.  Assuming that the crossover to
viscous dominating times signals the end of the growth of the gradients, 
one can estimate that 
\begin{equation}
(\nabla_i u_j)_{k} \sim \frac{u^*}{L^*}  (R^*)^{1/(2\alpha-1)}\mbox{,}
\end{equation}
which depends on $\alpha$.
This estimate serves for both strains and vorticity. 
The crossover acceleration then scales as the maximum value that the inertial term takes
\begin{equation}
a_k \sim \frac{(u^*)^2}{L^*} (R^*)^{(\alpha-2)/(2\alpha-1)}\mbox{.}
\end{equation}
This leads to a prediction that the index $\alpha$ effects the maximum acceleration in a particular
way.  In the case $\alpha<2$, the crossover acceleration $a_k$ shrinks with increasing Reynolds number,
while for $\alpha>2$, the crossover acceleration grows with acceleration.  Presumably the viscosity
caps the maximum acceleration, but these values still grow without bound with increasing Reynolds number. 
Note for all of these estimates $\alpha=1/2$ corresponding to the Leray solution \cite{leray} yields
an uncapped singularity.

Another way that the maximum gradients may be capped comes about due to the outer solution.  As these
events must live in a finite domain, and must be patched onto realistic boundary conditions, the
ideal form cannot exist out to large radius.  Thus either through this departure in the far field,
or by the outer condition deviating from the ideal form, the inner solution may miss the maximally
focused condition before viscosity would cap the event.

\section{Experimental observations}

\begin{figure}[h]
	\begin{center}
		\includegraphics[width=3.5in]{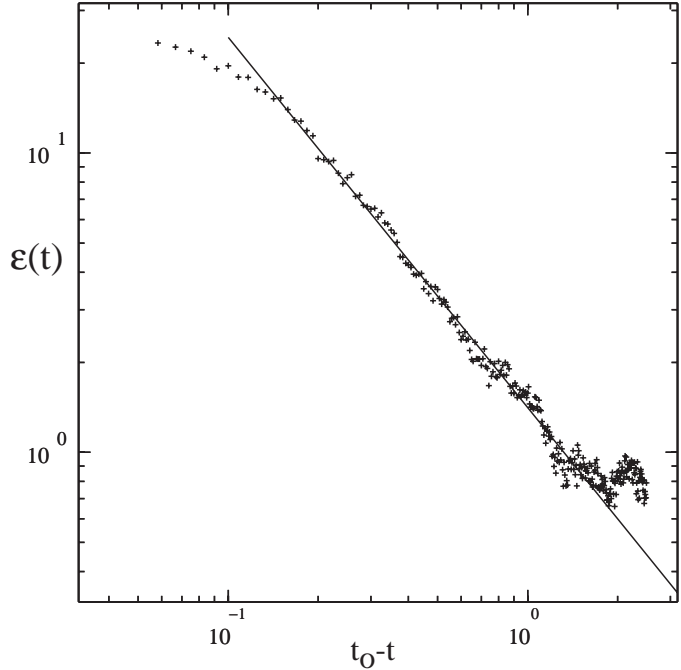}
	\end{center}
	\caption{The rise in the dissipation for the event shown in Fig. 1 has a power law form for some range in times.  Many such time series are examined to obtain events when the rise is of power law form for at least one decade in time, with a coefficient of regression for this decade of at least $R=0.98$.}
	\label{fig:fig2}
\end{figure}

Although the main parts of this paper are analytical, with 
some turbulent gradient data in hand,
I present some limited support for these ideas.  
A novel optical instrument has been utilized to
measure the gradients $\partial_i u_j(t)$ at the Kolmogorov scale (see \cite{zeff} for details about the
flow and instrument).  Focusing on the most extreme gradient events, I presume that one needs
to be both looking at the center of the event and near the critical time in order to have extremely
large gradients.  Since the measurement is Eulerian, solutions to Eq. 7 will both develop in place, and
more probably, be swept by the measurement volume.  As these are swept by, the time dependence
reflects the spatial dependence of the field; near the critical time this is the far field
dependence.

The far field has velocities scaling as $v_i \sim r^{(\alpha-1)/\alpha}$ so that gradients
behave as $\partial_i u_j \sim r^{-1/\alpha}$.  The dissipation per unit mass $\epsilon = (\nu/2)
\sum_{ij} (\partial_i u_j + \partial_j u_i)^2 \sim r^{-2/\alpha}$.  Fig. 1 shows one example
of a time trace for a large dissipation event.  As events are swept by one might occasion to see
$\epsilon(t) \sim t^{-2/\alpha}$.  
An extensive data set of gradients has been examined for events
where the dissipation behaves as 
\begin{equation}
\epsilon(t) \sim (t_\circ -t)^\delta
\label{scale}
\end{equation}
for some $t_1 < t < t_2 < t_\circ$.  So that they are at least minimally serious scaling I
require at least one decade of scaling, i.e. $(t_1-t_\circ)/(t_2-t_\circ)=10$ with a coefficient
of regression for a fit to Eq. \ref{scale}, $R^2>0.98$.  The event shown in Fig. 1 is redrawn
around the large dissipation growth showing (Fig. 2) one example of such a scaling.  Figure 3 shows a
histogram of the $\alpha = -2/\delta$ values for all 892 events observed of this type.  
Note that the
Kolmogorov value $\alpha=3/2$ occurs near the observed maxima.  
These observations are
only intended to partially support the analysis discussed as the bulk of this paper.  Further
experimental testing will be discussed elsewhere so as to not distract from the main points
presented here.

\begin{figure}[h]
	\begin{center}
		\includegraphics[width=3.5in]{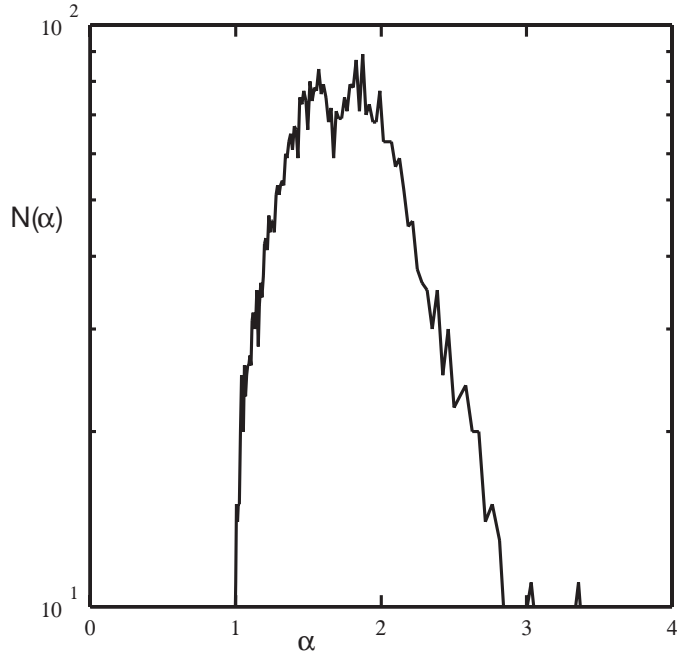}
	\end{center}
	\caption{Histogram of the observed values of the exponent $\alpha$, measured from the rise in dissipation, conditioned as indicated in Fig. 2.}
	\label{fig:fig3}
\end{figure}

\section{Conclusions}

The observation of events with $1 <= \alpha < 3$ may be associated with
the much prior commented on  multiscaling of the velocity field
\cite{sreeni}.  The solutions corresponding to K41 statistics is $\alpha = 3/2$
Why these solutions are dominant may be of
similar reasons to the Burgers case, where the same exponent also dominates.
One might hypothesize that locations where the gradients obey $M+M^2=\Pi_2$ and
the second derivatives ($\partial_i \partial_j v_k$) vanish may be the
locations susceptible to this types of blowup -- in analogy to
inviscid Burgers shock development.  The dominance then of the cubic terms
in the velocity near field leads, through a characteristic collapse, to
$\alpha = 3/2$ events.  While other types of $\alpha$ events also 
appear possible, a selection mechanism is not obvious.

And then one must deal with the fact that these solutions are self-similar
around one point in time and space, while turbulence is a statistically
self-affine (or multi-affine) phenomena\cite{meneveau,sreeni,sreeni2}.  What is needed is a number
of events associated with Eq. \ref{simeq} occurring at many spatial
locations and times.  This may be accommodated if these solutions give
rise to the growth of additional events nearby due to the form the 
intermediate field takes.  In particular, if a sufficiently complicated
angular structure for the far field occurs, when that must flow into
the relatively simple near field,  there may be locations with
the right type of structure (perhaps in the second order derivative
field) which gives rise to new growing collapse events.

Finally, these events are distinct from long lived vortices others have observed,
but perhaps causally connected.  A sudden growth of the gradients around a point
might be expected when two tube-like vortices collide obliquely \cite{kerrpriv}.
Also, I acknowledge that more complicated solutions, 
perhaps with less trivial length scaling may also play an important role.

{\noindent {\bf Acknowledgements } }

I would like to greatfully acknowledge the support of the National 
Science Foundation and the Research Corporation, and stimulating advice from 
T. Antonsen, J. Gibbon, R.M. Kerr, C.D. Levermore, J. Lister, E. Ott, 
R. Pego, K.R. Sreenivasan, and E. Tadmor. The experiments could not have
been conducted without the efforts and advice of B.W. Zeff, D.D. Lanterman, 
R. McAllister, R. Roy, and E.J. Kostelich.

\end{document}